\documentclass[10pt,preprint,pre,floats,english,superscriptaddress]{revtex4-1}
\usepackage{amsthm}
\usepackage{epsf}
\usepackage[scanall]{psfrag}
\usepackage{dsfont}
\usepackage[T1]{fontenc}
\usepackage[latin9]{inputenc}
\usepackage{lmodern}
\usepackage{geometry}
\usepackage{caption,subcaption}
\usepackage{amsmath}
\usepackage{amssymb}
\usepackage{graphicx}
\usepackage{color}
\usepackage{chngpage}

\usepackage{multirow}

\makeatletter
\@ifundefined{textcolor}{}
{%
 \definecolor{BLACK}{gray}{0}
 \definecolor{WHITE}{gray}{1}
 \definecolor{RED}{rgb}{1,0,0}
 \definecolor{GREEN}{rgb}{0,1,0}
 \definecolor{BLUE}{rgb}{0,0,1}
 \definecolor{CYAN}{cmyk}{1,0,0,0}
 \definecolor{MAGENTA}{cmyk}{0,1,0,0}
 \definecolor{YELLOW}{cmyk}{0,0,1,0}
 }

\makeatother

\usepackage{babel}

\begin{document}
\begin{flushright}
DAMTP-2013-67
\end{flushright}

\title{Potential Energy Landscapes for the 2D XY Model: \\
Minima, Transition States and Pathways}

\author{Dhagash Mehta}
\email{dbmehta@ncsu.edu}
\affiliation{Dept of Mathematics, North Carolina State University, Raleigh, NC 27695, USA.}

\author{Ciaran Hughes}
\email{ch558@cam.ac.uk}
\address{The Department of Applied Mathematics and Theoretical Physics, The
University of Cambridge, Clarkson Road, Cambridge CB3 0EH, UK.}

\author{Mario Schr\"ock}
\email{mario.schroeck@uni-graz.at}
\affiliation{Institut f\"ur Physik, FB Theoretische Physik, Universit\"at Graz, 8010 Graz, Austria.}

\author{David J.~Wales}
\email{dw34@cam.ac.uk}
\affiliation{University Chemical Laboratories, Lensfield Road, Cambridge CB2 1EW, UK.}

\begin{abstract}
We describe a numerical study of the potential energy landscape for the two-dimensional XY
model (with no disorder), considering
up to $100$ spins and CPU and GPU implementations of local optimization,
focusing on minima and saddles of index one (transition states).
We examine both periodic and anti-periodic boundary conditions, and show that the
number of stationary points located
increases exponentially with increasing lattice size.
The corresponding disconnectivity graphs exhibit funneled landscapes;
the global minima are readily located
because they exhibit relatively large basins of attraction compared to the higher energy minima
as the lattice size increases.
\end{abstract}
\maketitle

\section{Introduction}
\label{sec:intro}

Stationary points of a potential energy function, defined as configurations where the
gradient vanishes, play a key
role in determining many phenomena in physics and chemistry.
An extensive framework of conceptual and computational tools has been developed
corresponding to the potential energy landscape approach
\cite{Wales:04,Wales05,WalesB06,Wales12,RevModPhys.80.167}, with applications
to many-body systems as diverse as metallic clusters,
biomolecules, structural glass formers, and coarse-grained models of soft
and condensed matter.
In all these examples, stationary points of
a high-dimensional potential energy
function are considered. Due to the non-linear nature of the potential energy
as a function of coordinates arising
in most real world applications,
obtaining the stationary points analytically is not feasible. Hence, one has to
rely on numerical methods to obtain the necessary information.

In the present contribution we initiate an extensive numerical analysis of the stationary points of a
well-known example, the XY model (without any disorder),
for a Hamiltonian defined in terms of the potential energy:
\begin{equation}
H=\frac{1}{N^d}\sum_{j=1}^{d}\sum_{\textbf{i}}[1- \cos(\theta_{\textbf{i}+\hat{\boldsymbol\mu}_j}-\theta_{\textbf{i}})],\label{eq:F_phi},
\end{equation}
where $d$ is the dimension of a lattice, $\hat{\boldsymbol\mu}_j$ is the $d$-dimensional unit vector in the $j$-th direction,
i.e.~$\hat{\boldsymbol\mu}_1=(1,0,\ldots,0)$, $\hat{\boldsymbol\mu}_2=(0,1,0,\ldots,0)$, etc.,
$\textbf{i}$ stands for the
lattice coordinate $(i_{1},\dots,i_{d})$, and the sum over ${\textbf{i}}$ represents a sum over all $i_{1},\dots,i_{d}$ each
running from $1$ to $N$, and each $\theta_{\textbf{i}}\in(-\pi,\pi]$.
Hence $d$ is the dimension of the lattice, and $N$ is the number of sites for each dimension,
so the number of $\theta$ values required to specify the configuration is $N^d$.
The boundary conditions are given by $\theta_{\textbf{i}+N\hat{\boldsymbol\mu}_j}=(-1)^{k}\theta_{\textbf{i}}$ for
$1\le j\le d$,
where $N$ is the total number of lattice sites in each dimension,
with $k=0$ for periodic boundary conditions (PBC) and $k=1$ for
anti-periodic boundary conditions (APBC). With PBC there is a global
degree of freedom leading to a one-parameter family of solutions,
as all the equations are unchanged under $\theta_{\textbf{i}}\to\theta_{\textbf{i}}+\alpha,\forall \textbf{i}$,
where $\alpha$ is an arbitrary constant angle,
reflecting the fact that the model has global O($2$) symmetry. We remove this
degree of freedom by fixing one of the variables to zero: $\theta_{(N,N,\ldots,N)}=0$.
We have included the factor $1/N^d$ to facilitate comparisons between systems of different sizes.
In the present contribution we mainly focus on analysis of local minima and the
pathways between them that are mediated by
transition states (saddles of index one, with a single negative Hessian eigenvalue \cite{murrelll68}).
% for the 2D XY model.

$H$ appears in many different
contexts: first, in statistical physics $H$
is known as the XY Model Hamiltonian and is known to exhibit a Kosterlitz-Thouless
transition \cite{kosterlitz1973ordering}. It describes a system of $N$ classical
planar spin variables, where each spin is coupled to its four nearest
neighbors on the lattice.
This representation is
employed in studies of low temperature superconductivity,
superfluid helium, hexatic liquid crystals, and Josephson junction arrays.
$H$ also corresponds to the lattice Landau gauge functional for a compact
$U(1)$ lattice gauge theory \cite{Maas:2011se,Mehta:2009,Mehta:2010pe},
and to the nearest-neighbor Kuramoto model with homogeneous
frequency, where the stationary points constitute special configurations
in phase space from a non-linear dynamical systems viewpoint
\cite{acebron2005kuramoto}.

The XY model is among the simplest lattice spin models in which an
energy landscape approach based on stationary points of the Hamiltonian
in a continuous configuration space
is appropriate (unlike, for example, the Ising model whose configuration
space is discrete).
Nevertheless, we find that the potential energy landscape supports a wide range of
interesting features, and proves to be very helpful in analyzing
the characteristic structure, dynamics, and thermodynamics.

%The mean-field XY model was solved analytically in
%\cite{Casetti:June2003:0022-4715:1091,PhysRevLett.82.4160} and the one-dimensional
%XY model was studied in
%Ref.~\cite{Casetti:June2003:0022-4715:1091,PhysRevLett.82.4160}, where a class
%of stationary points was identified analytically.

In Ref.~\cite{Mehta:2009,Mehta:2010pe} all
the stationary points of the one-dimensional XY model were found, including interesting
classes, such as stationary wave solutions. In Ref.~\cite{Mehta:2009,vonSmekal:2007ns,vonSmekal:2008es},
all the stationary points for the one-dimensional model with anti-periodic
boundary conditions were characterized. Solving the stationary equations for the XY model
in more than one dimension turned out to be a difficult task and has
not been completed to date. In Ref.~\cite{Mehta:2009}
it was shown how the stationary equations could be
viewed as a system of polynomial equations, and the
numerical polynomial homotopy continuation (NPHC) method was employed to
find all the stationary points for small lattices in two dimensions.
This method was subsequently used to study the potential
energy landscapes of various other models in statistical mechanics and particle
physics \cite{Mehta:2011xs,Mehta:2011wj,Maniatis:2012ex,Kastner:2011zz,Mehta:2012wk,PhysRevE.85.061103,PhysRevE.87.052143,PhysRevD.88.026005,MartinezPedrera:2012rs,He:2013yk}.
In particular, it was employed to study the potential energy landscape
of the two-dimensional (2D) XY model \cite{Mehta:2009zv,Hughes:2012hg,Casetti-Mehta:2012}.
In the latter work, along with all the isolated solutions
for a small $3\times 3$ lattice, two types of singular solutions were
characterized: (1) isolated singular solutions, where the Hessian matrix is singular
(these solutions are in fact multiple solutions); and (2) a
continuous family of singular solutions.
It was shown that one can construct one-, two-, etc. parameter solutions, even after fixing
the global $O(2)$ freedom.

In Ref.~\cite{Hughes:2012hg} application
of the conjugate gradient method for small $N$ suggested
that the number of local minima for the 2D XY model would increase exponentially.
One of the important results of the current paper is to verify this conjecture and make it
more precise by characterizing the landscapes for larger $N$
values, while improving on the earlier results for the number of minima.

All the above-mentioned minimization methods have a common shortcoming because
they cannot deal with even moderately high $N$ (of the order $100$ angles).
Finding saddles is even harder, and so far the only available results are
for $3\times 3$ lattices. In the current paper, we use two more powerful
tools to explore the potential energy landscape (PEL)
 of the 2D XY model in a detailed manner, namely
the {\tt OPTIM}
package \cite{optim}, and a GPU-implementation of the overrelaxation
method. These approaches can explore the PELs of the 2D XY model with $100$ spins
and beyond. In the next two Sections, we describe the functionalities of {\tt OPTIM}
that we have used in our work and the GPU implementation of the overrelaxation
method.

\section{Methods}

\subsection{Geometry Optimization}
\label{sec:geopt}

The {\tt OPTIM} program includes a wide variety of geometry optimization tools
for characterizing stationary points and the pathways that connect them
\cite{optim}.
The most efficient \cite{AsenjoSWF13} gradient-only minimization algorithm implemented in
{\tt OPTIM} is a modified version of the limited-memory
Broyden--Fletcher--Goldfarb--Shanno (LBFGS) algorithm \cite{Nocedal80,lbfgs}.
Both gradient-only and second derivative-based eigenvector-following \cite{Wales92,Wales93d}
and hybrid eigenvector-following algorithms \cite{munrow99,kumedamw01}
are available for single- and double-ended \cite{TrygubenkoW04} transition state searches.
Stationary points with any specified Hessian index can also be located \cite{2003JChPh.11912409W}.

We used {\tt OPTIM} to sample minima and transition states for the 2D XY model with
both PBC and APBC, using exclusively single-ended search algorithms.
In particular, we refined $500,000$ random initial guesses for all lattice sizes up to $N=10$,
i.e., a total of $100$ spins.
In each case there also exist degenerate stationary points related by symmetry
operations of the Hamiltonian with
$\theta_{\textbf{i}} \rightarrow -\theta_{\textbf{i}}$
for all $N^d$ angular variables as well as $\theta_{\textbf{i}}
\rightarrow \theta_{\textbf{i}} \pm (\pi, \pi, \dots, \pi)$.

\subsection{GPU Implementation of Overrelaxation}
\label{sec:GPU}

Another approach that we applied to obtain as many minima as
possible employed the (over)relaxation algorithm
exploiting graphics processing units (GPUs), which offer a high level
of parallelism and thus enabled us to generate large samples within
a practical amount of computer time.
The idea of the relaxation algorithm is to sweep over the lattice while optimizing
the Hamiltonian locally on each lattice site.
Our implementation is based on the \emph{www.cuLGT.com} code \cite{Schrock:2012fj}.

In practice we employed four cards of the NVIDIA Tesla C2070 and launched 1024 thread blocks
(i.e.~samples) per GPU. We kept the grid size as 1024 blocks per GPU fixed and then
cycled over
$2^{17}=131072$ iterations, resulting in around $0.134$ billion samples per lattice size.
We set the overrelaxation parameter to 1.0, i.e.~standard relaxation, to increase
the chance of finding minima with small basins of attraction.
For each sample we stored the value of the minimum to which the relaxation algorithm
converged along with the $N^d$ corresponding $\theta$-coordinates,
and subsequently sorted these values via bitonic sort \cite{Bat68},
again accelerated by the GPU. As a stopping criterion we required
the largest gradient over all lattice sites to be smaller than $10^{-12}$ (reduced units).
The whole simulation was performed in double precision to reduce numerical inaccuracies.

% DJW up to here

\section{Results}
\label{sec:results}

We first point out that the global minimum of this model, as it does not have any disorder,
is well known: for the APBC case, $\theta_{\textbf{i}}=0$ or $\pi$
for all $N^d$ angular variables
are the two global minima of the model at which $H=0$.
Similarly, for the PBC case, because we have fixed the $\mathcal{O}(2)$ symmetry,
the unique global minimum corresponds to $\theta_{\textbf{i}}=0$ for all angles, again with $H=0$.

\subsection{Number of Minima and Transition States}
\label{sec:number}

In Table \ref{tab:minima-transition-states} we summarize the number of minima and
transitions states located for each $N$. Here, in addition to finding minima
and transition states for larger lattices, we have also improved our previous
results for the number of minima from Ref.~\cite{Hughes:2012hg} at smaller
$N$ with the help of the more powerful algorithms.
Saddles of index one were only obtained from the {\tt OPTIM} runs. Since the 2D
XY model possesses a number of discrete symmetries, as discussed in
\cite{Hughes:2012hg}, we also tabulate the number of distinct minima and
transition states in the table,
i.e.~solutions unrelated by symmetries of the Hamiltonian.
In contrast to $N\leq 8$, for larger $N$ our samples will be substantially incomplete,
even though we have
found around $5.5$ million minima for the $N=16$ case. As expected from symmetries of the Hamiltonian \cite{Hughes:2012hg},
for each $N$ with PBC the global minimum is unique, the next minimum is 4-fold
degenerate, then $N^2/2$-fold degenerate (if $N$ is even), then
$2 N^2$, $4 N^2$, $2 N^2$,...., $N^2/2$ (if $N$ is even), and the highest
energy minimum is 4-fold degenerate.
For the APBC case the global minimum is $2$-fold degenerate, and all other minima are at least $N^2$ degenerate.

\begin{figure}[htbp]
\psfrag{N}[tc][tc]{$N$}
\psfrag{Number of Minima}[bc][bc]{Number of Minima}
\psfrag{10}[cr][cr]{10}
\psfrag{0}[bl][bl]{\small 0}
\psfrag{1}[bl][bl]{\small 1}
\psfrag{2}[bl][bl]{\small 2}
\psfrag{3}[bl][bl]{\small 3}
\psfrag{4}[bl][bl]{\small 4}
\psfrag{5}[bl][bl]{\small 5}
\psfrag{6}[bl][bl]{\small 6}
\psfrag{7}[bl][bl]{\small 7}
\psfrag{ 4}[tc][tc]{4}
\psfrag{ 6}[tc][tc]{6}
\psfrag{ 8}[tc][tc]{8}
\psfrag{ 10}[tc][tc]{10}
\psfrag{ 12}[tc][tc]{12}
\psfrag{ 14}[tc][tc]{14}
\psfrag{ 16}[tc][tc]{16}
\psfrag{all minima (PBC)}[cr][cr]{all minima (PBC)}
\psfrag{dist. minima (PBC)}[cr][cr]{distinct minima (PBC)}
\psfrag{all saddles idx. 1 (PBC)}[cr][cr]{all index 1 saddles (PBC)}
\psfrag{dist. saddles idx. 1 (PBC)}[cr][cr]{distinct index 1 saddles (PBC)}
\psfrag{all minima (APBC)}[cr][cr]{all minima (APBC)}
\psfrag{dist. minima (APBC)}[cr][cr]{distinct minima (APBC)}
\psfrag{all saddles idx. 1 (APBC)}[cr][cr]{all index 1 saddles (APBC)}
\psfrag{dist. saddles idx. 1 (APBC)}[cr][cr]{distinct index 1 saddles (APBC)}
\centerline{\includegraphics[width=0.99\textwidth]{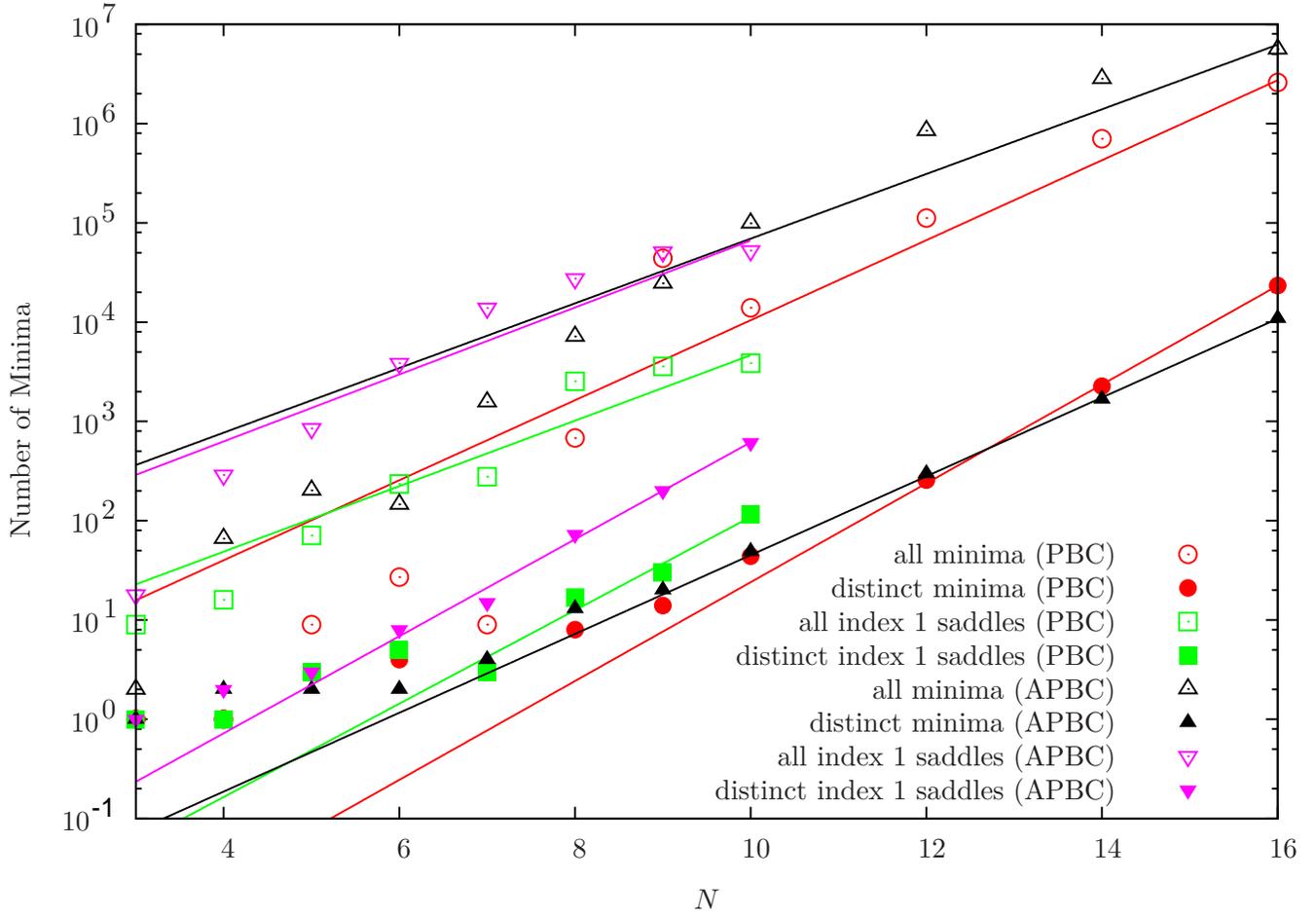}}
\caption{Number of minima as a function of the number of lattice sites, $N$, for each dimension.
The straight lines are the corresponding best fits for the data-points, i.e.~the number of distinct minima and the total number of minima
including degeneracies increases roughly exponentially with increasing $N$.}
\label{fig:plot_N_vs_min_ts}
\end{figure}

The number of minima and transition states located as a function
of $N$ are plotted in Figure \ref{fig:plot_N_vs_min_ts}. The plot clearly
shows that the total number of minima (including degeneracies),
the number of distinct minima, the
total number transition states, and the number of distinct transition states, all
increase roughly exponentially with increasing $N$,
as expected \cite{stillingerw84,2003JChPh.11912409W}.
We also observe abrupt jumps at
$N=7$ and $9$ for the PBC case, and at $N=6$ for the APBC cases in this plot,
though the precise reason of this behavior is not clear.
It is possible that the jumps are caused by sampling issues,
but there could be
a more subtle explanation; for example, certain lattices for particular values of $N$ may
restrict the possible classes of minima.

For both APBC and PBC the energies of the local minima
shift towards lower energy as $N$ increases,
tending to accumulate near the global
minima. This behavior has previously been observed in the 1D XY model with PBC
\cite{Hughes:2012hg}.
In this case, the potential energy distribution of the minima has a spike for the
global minimum at $H = 0$ and a two-fold degeneracy for other minima in $H \in (0, 1]$.
Since every minimum in the ordered 1D XY model has a higher-dimensional
analogue with the same energy \cite{Hughes:2012hg}, it is not
surprising that the 2D PBC case exhibits similar behavior.
Straightforward construction of higher-dimensional saddles from lower-dimensional
ones is not possible in APBC, so it is interesting to
see the 2D APBC XY model behaving differently from the 1D APBC
XY model, but similar to the 2D PBC case. In future work, we also intend to \textit{certify} these solutions
using techniques based on Smale's $\alpha$-theorem \cite{Mehta:certification2013}.

\subsection{Disconnectivity Graphs}
\label{sec:discon}

Disconnectivity graphs have provided a particularly useful tool for
visualizing potential and free energy landscapes  \cite{beckerk97,walesmw98,KrivovK02,EvansW03} in systems ranging from
atomic and molecular clusters to soft and condensed matter and biomolecules \cite{Wales:04,Wales05,WalesB06}.
In particular, this construction enables the lowest potential or free energy
barriers to be faithfully represented, and can help us to understand how
observable properties emerge from features of the landscape \cite{Wales10a}.

To produce a disconnectivity graph we require a kinetic transition network \cite{NoeF08,pradag09,Wales10a},
which can be defined by a database of local minima and the transition states that connect them \cite{beckerk97}.
We then choose a regular energy spacing, $\Delta V$, and determine how the minima are partitioned into subsets
(superbasins \cite{beckerk97}) at energies $V_0,\ V_0+\Delta V,\ V_0+2\Delta V,\ldots$.
These subsets consist of minima that can interconvert via index one saddles
that lie below the energy threshold.
For a high enough threshold all the minima can interconvert and there is just one superbasin, unless there
are infinitely high barriers.
As the threshold energy decreases the superbasins split apart, and
this splitting is represented in the disconnectivity graph by lines
connecting subgroups to parent superbasins at the threshold energy above.
The superbasins terminate at the energies of individual local minima, which may
be grouped together for degenerate states related by symmetry operations of the Hamiltonian.

The significance of the disconnectivity graph construction stems primarily from
the insight it provides into the global thermodynamics and kinetics of the system
in question.
For example, if the landscape supports alternative morphologies separated by a
high barrier then we anticipate a separation of relaxation time scales and
associated features in the heat capacity \cite{Wales:04,Wales05,WalesB06,Wales10a}.
Several limiting cases have been identified for the organization of the landscape,
distinguishing good `structure seeking' systems, which exhibit efficient
relaxation to the global minimum, from models with glassy characteristics \cite{walesmw98}.
These visualisations have much in common with the `energy lid' and `energy threshold'
approaches of Sibani, Sch\"on, and coworkers \cite{SibaniSchoen93,SibaniSch94,schon96,schonpj96,schon02}.

\begin{figure}[hp]
\center
\psfrag{    0.00}[cr][cr]{0.0}
\psfrag{    0.10}[cr][cr]{0.1}
\psfrag{    0.20}[cr][cr]{0.2}
\psfrag{    0.30}[cr][cr]{0.3}
\psfrag{    0.40}[cr][cr]{0.4}
\psfrag{    0.50}[cr][cr]{0.5}
\psfrag{    0.60}[cr][cr]{0.6}
\psfrag{    0.70}[cr][cr]{0.7}
\psfrag{    0.80}[cr][cr]{0.8}
\psfrag{    0.90}[cr][cr]{0.9}
\psfrag{    1.00}[cr][cr]{1.0}
\includegraphics[width=0.33\textwidth]{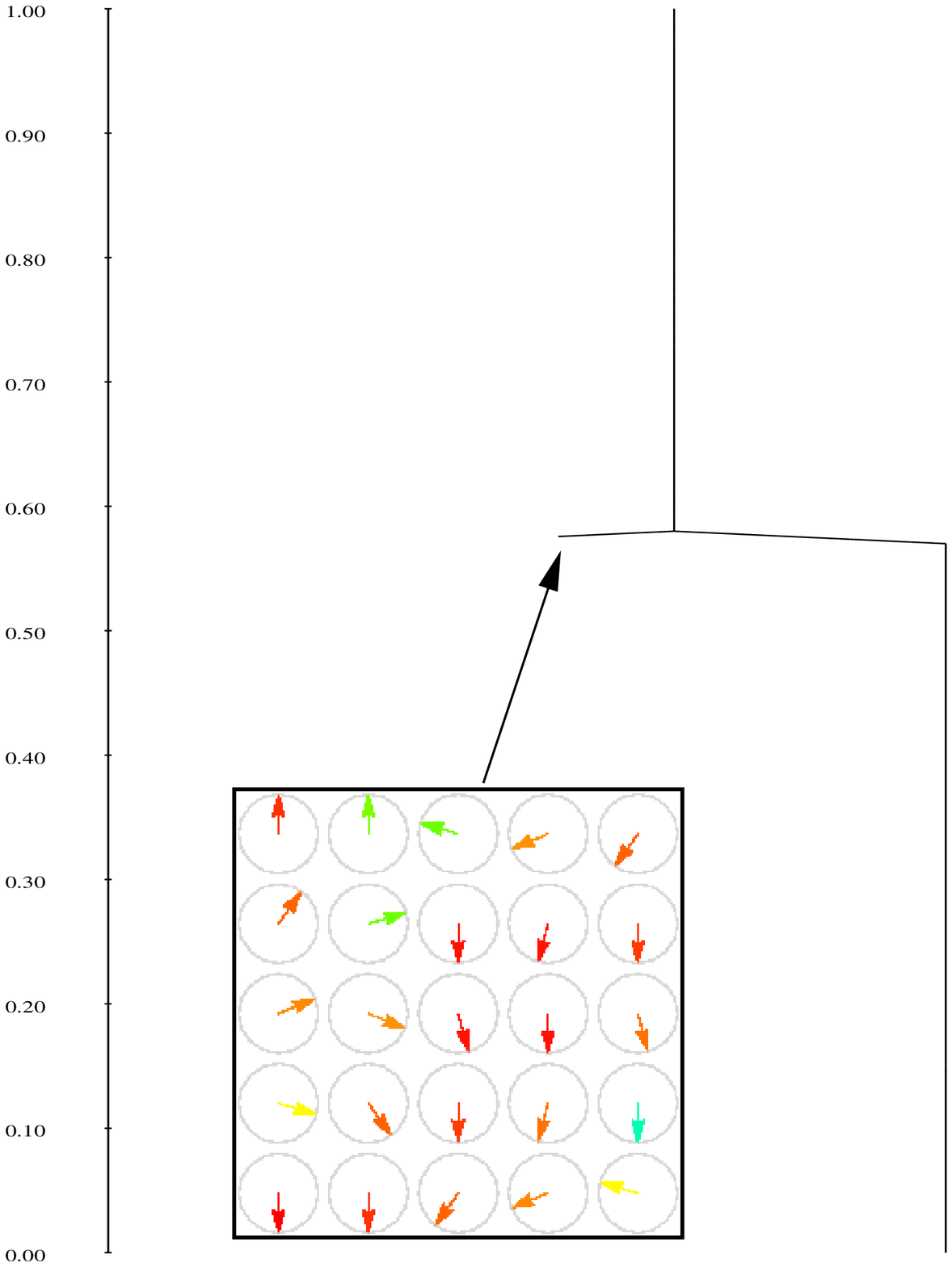}
\includegraphics[width=0.33\textwidth]{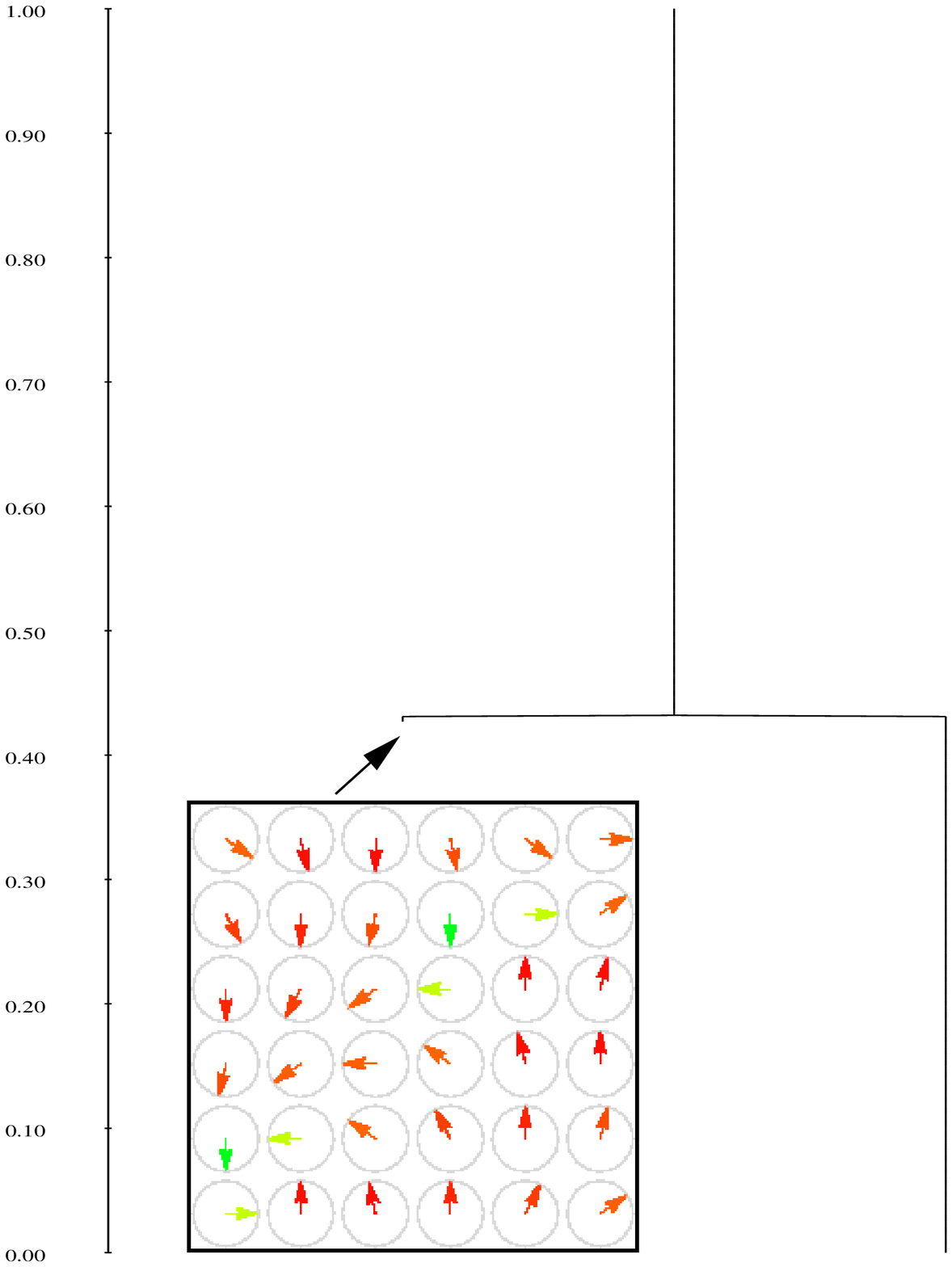}
\vskip 0.5 true cm
\includegraphics[width=0.33\textwidth]{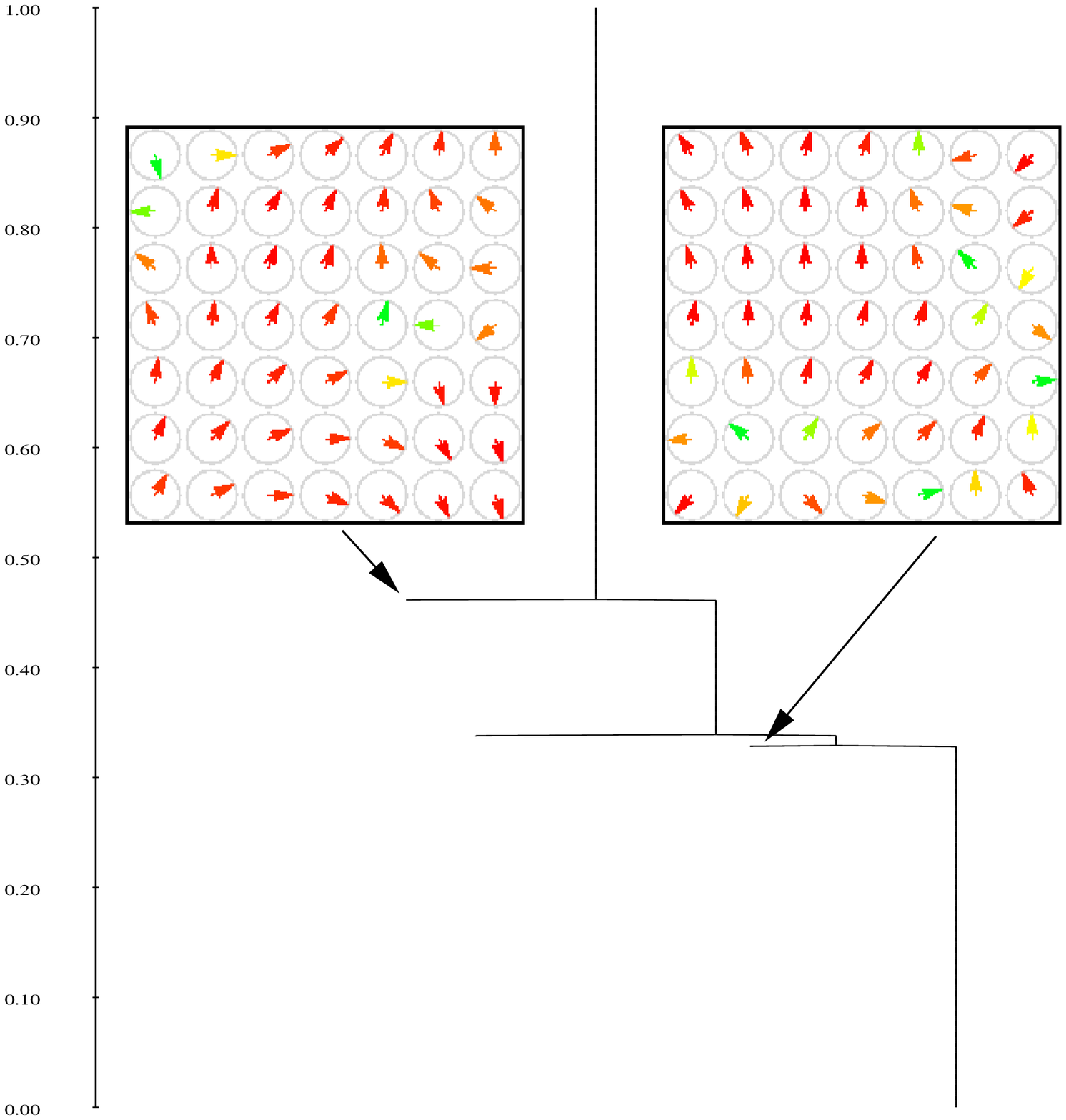}
\includegraphics[width=0.33\textwidth]{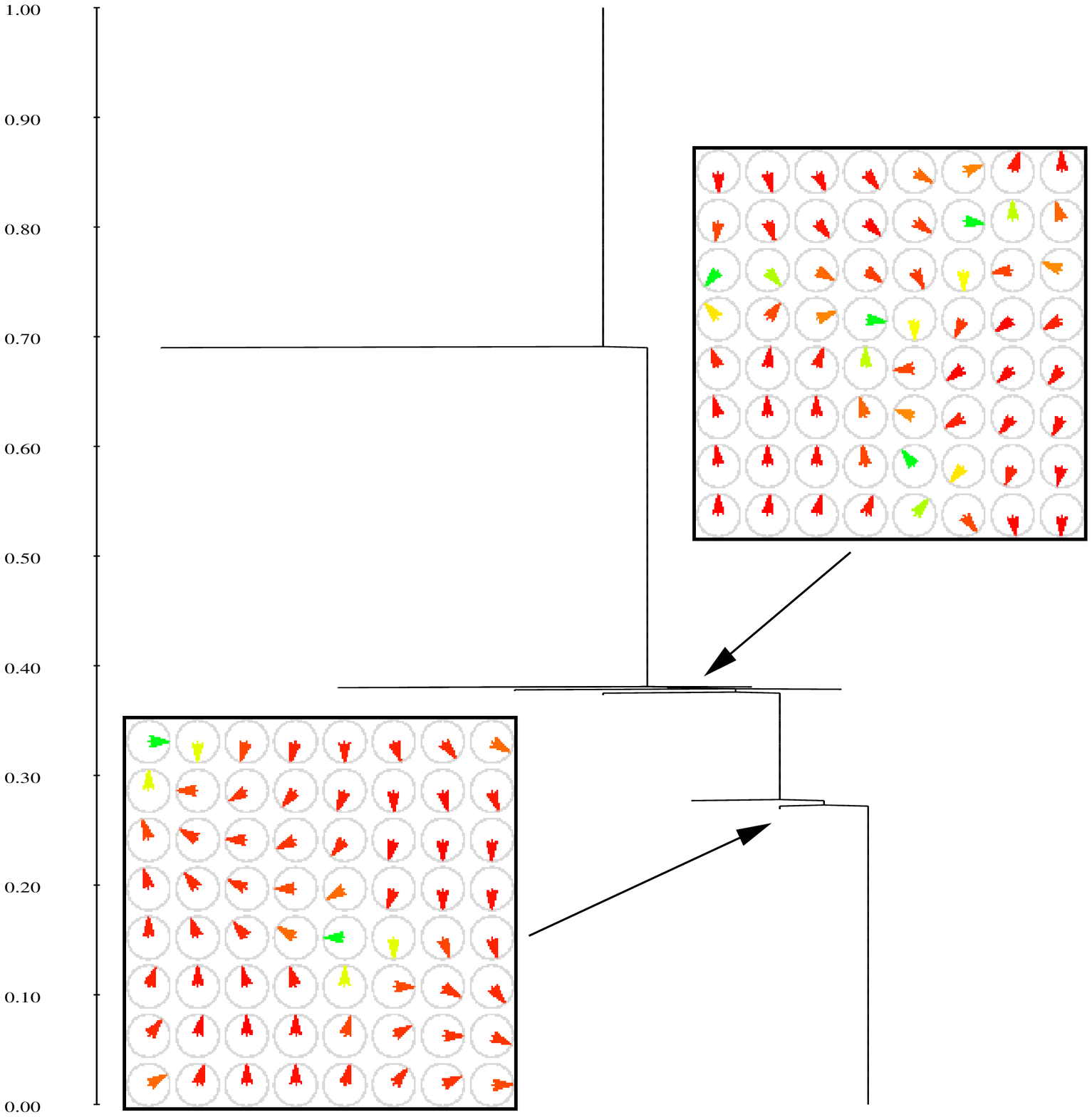}
\vskip 1.0 true cm
\includegraphics[width=0.33\textwidth]{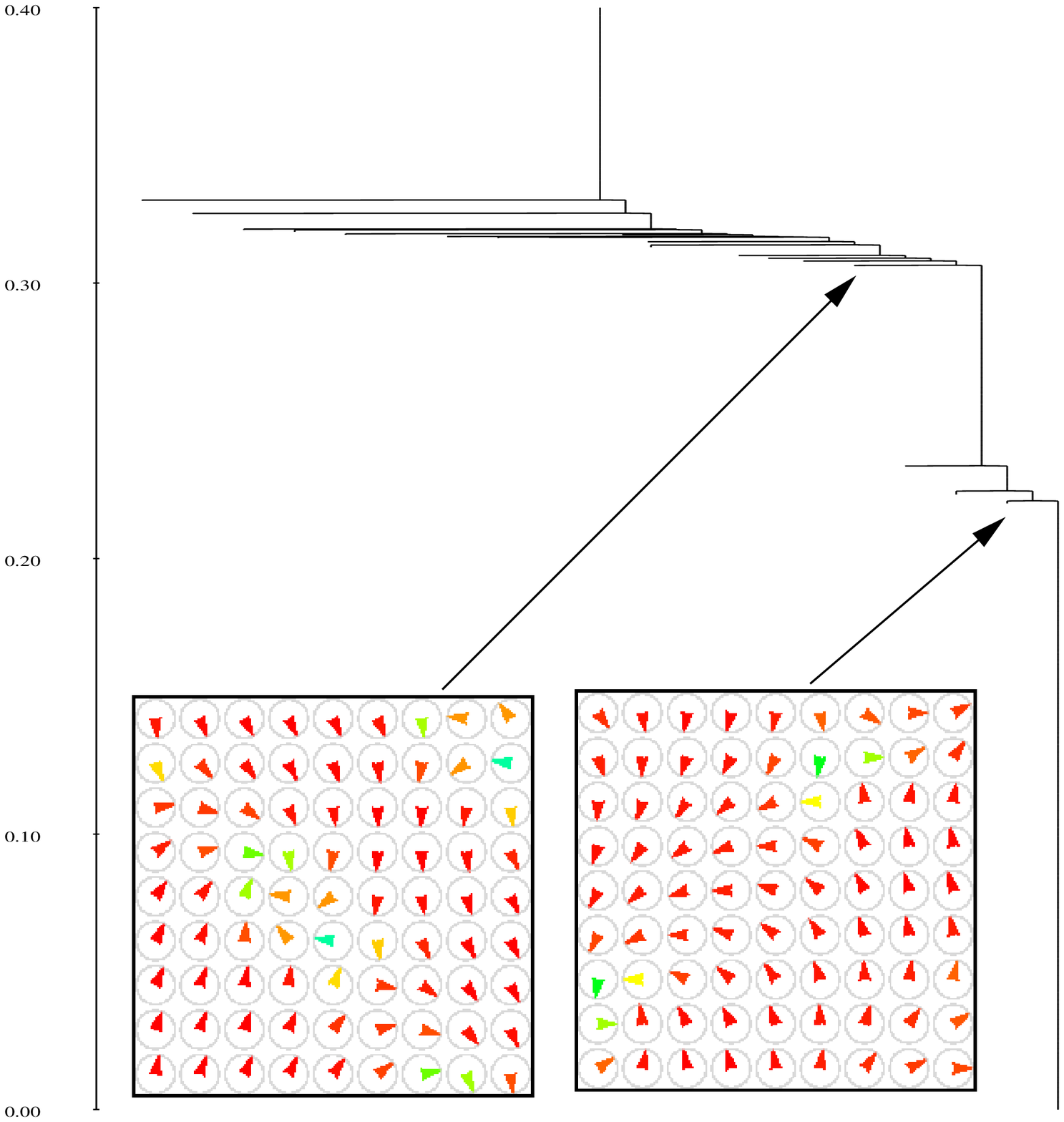}
\includegraphics[width=0.33\textwidth]{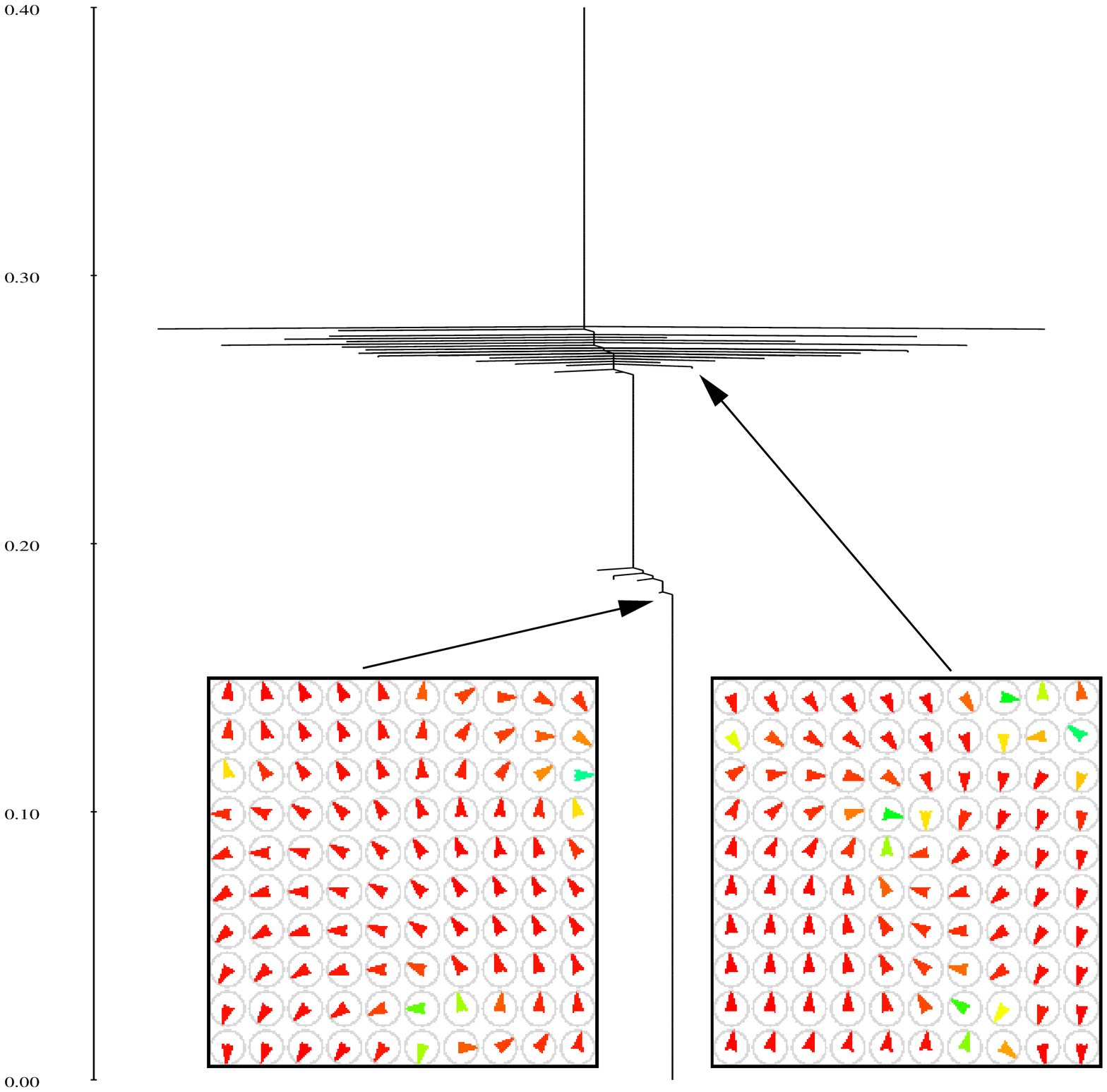}
\caption{Disconnectivity graphs for the $5\times 5$, $6\times 6$, $7\times 7$,
$8\times 8$, $9\times 9$ and $10\times 10$ APBC lattices. Each of the two insets
represents an example minimum for the corresponding
$N\times N$ lattice. Each arrow in these insets represents the corresponding value of
$\theta_{{\bf i}}$ at the lattice-site ${\bf i}$. At each lattice site ${\bf i}$, we compute the
local energy $\sum_{j= 1}^{d}(1- \cos(\theta_{\textbf{i}+\hat{\boldsymbol\mu}_j}-\theta_{\textbf{i}}))$,
which is in
the range $[0,4]$. We colour the arrows
red-orange-yellow-green-blue-indigo-violet from the lowest to highest
local energies.}
\label{fig:discon_APBC}
\end{figure}

\begin{figure}
\center
\psfrag{    0.00}[cr][cr]{$0.0$}
\psfrag{    0.10}[cr][cr]{}%{$0.1$}
\psfrag{    0.20}[cr][cr]{$0.2$}
\psfrag{    0.30}[cr][cr]{}%{$0.3$}
\psfrag{    0.40}[cr][cr]{$0.4$}
\psfrag{    0.50}[cr][cr]{}%{$0.5$}
\psfrag{    0.60}[cr][cr]{$0.6$}
\psfrag{    0.70}[cr][cr]{}%{$0.7$}
\psfrag{    0.80}[cr][cr]{$0.8$}
\psfrag{    0.90}[cr][cr]{}%{$0.9$}
\psfrag{    1.00}[cr][cr]{$1.0$}
\psfrag{    1.10}[cr][cr]{}%{$1.1$}
\psfrag{    1.20}[cr][cr]{$1.2$}
\psfrag{    1.30}[cr][cr]{}%{$1.3$}
\psfrag{    1.40}[cr][cr]{$1.4$}
\psfrag{    1.50}[cr][cr]{}%{$1.5$}
\includegraphics[width=0.33\textwidth]{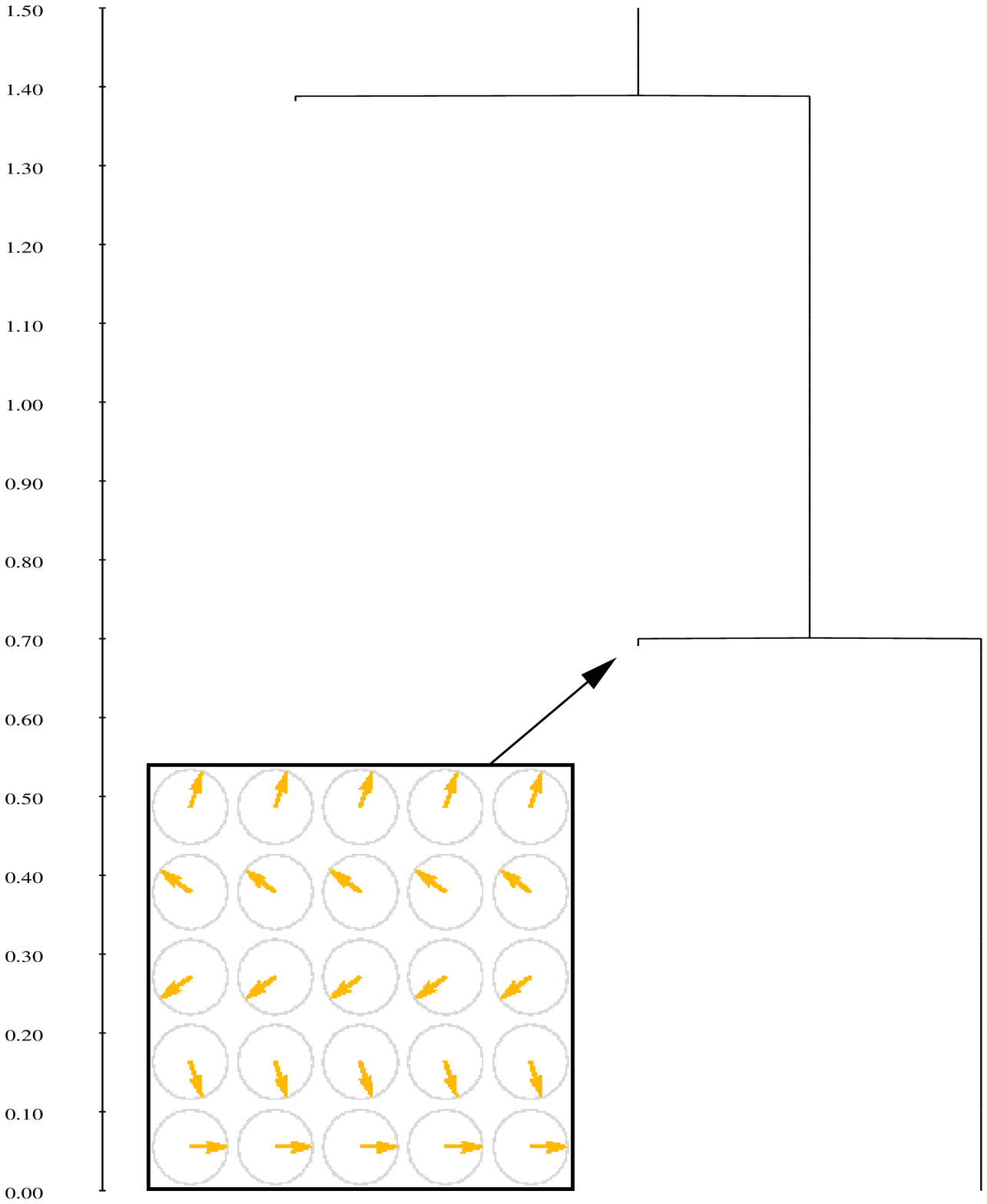}
\includegraphics[width=0.33\textwidth]{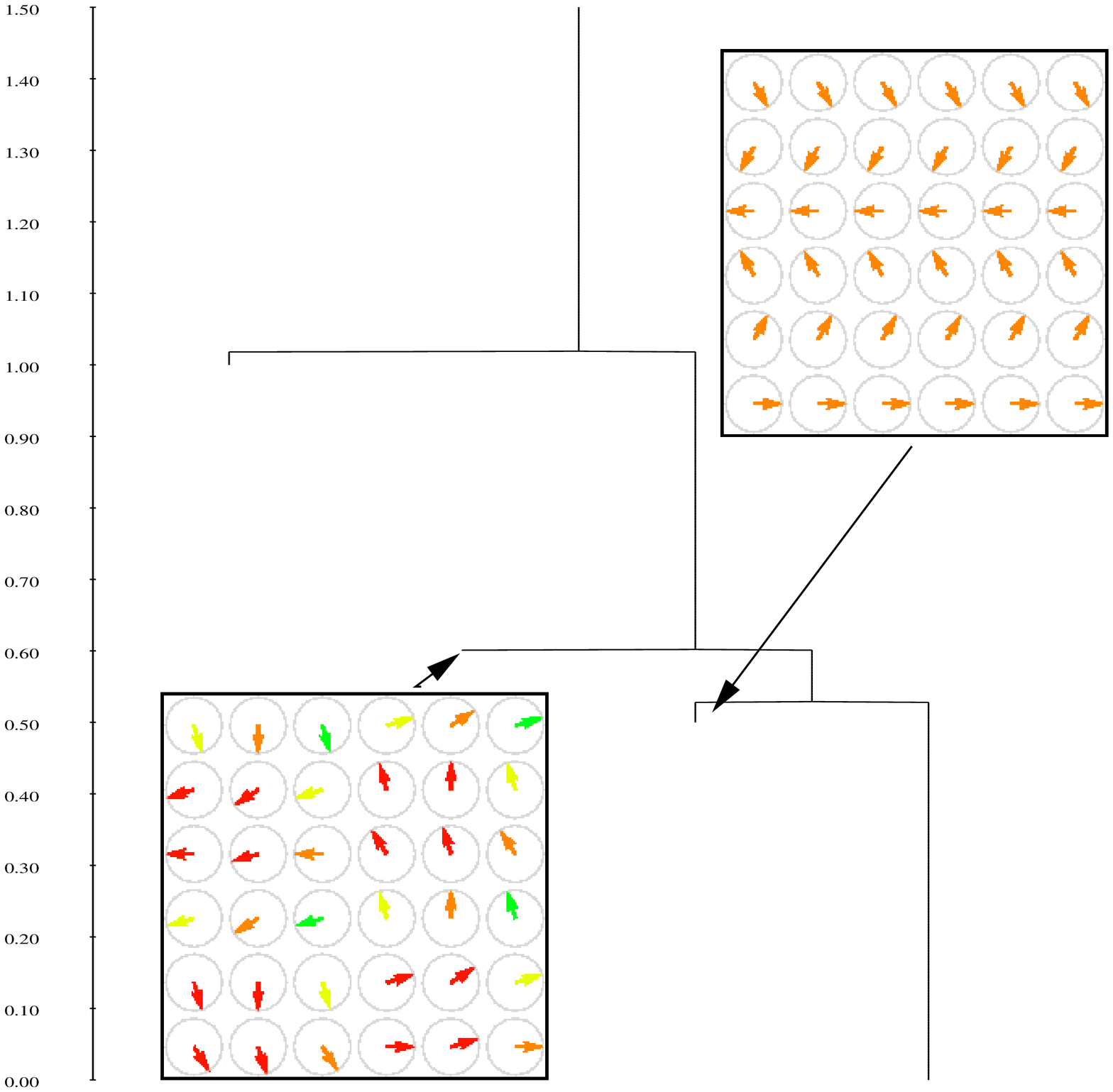}
\vskip 0.5 true cm
\includegraphics[width=0.33\textwidth]{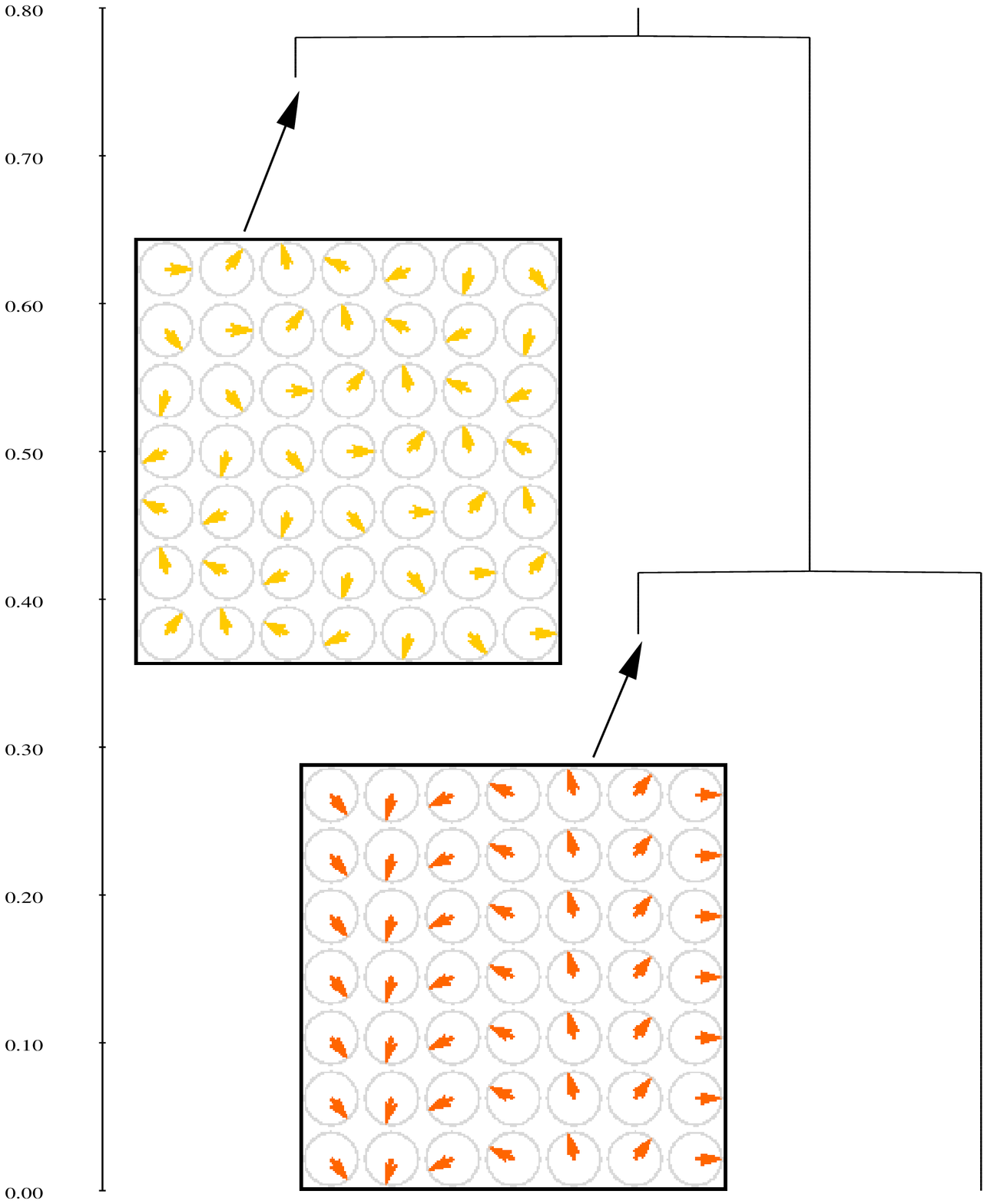}
\includegraphics[width=0.33\textwidth]{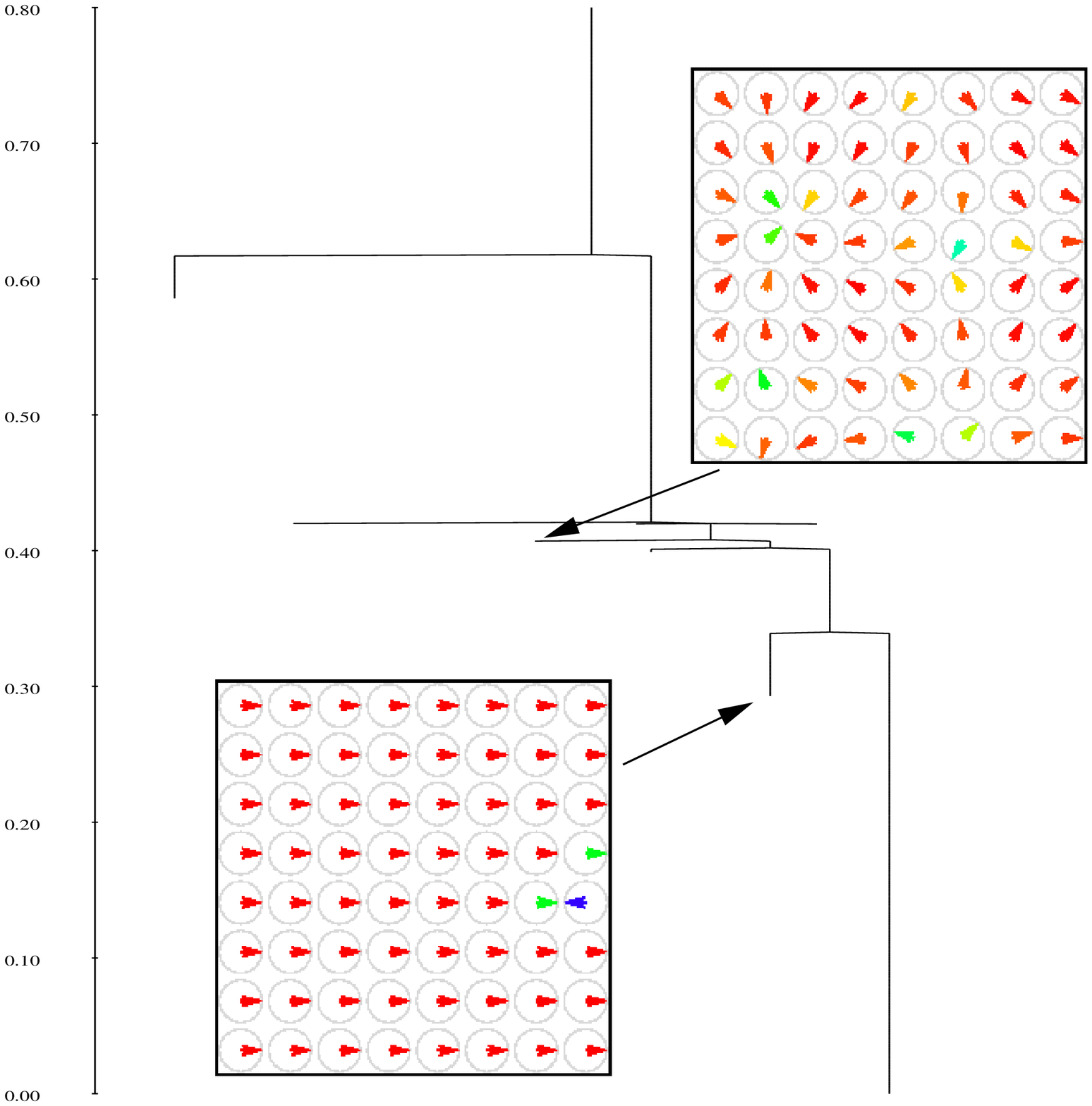}
\vskip 0.5 true cm
\includegraphics[width=0.33\textwidth]{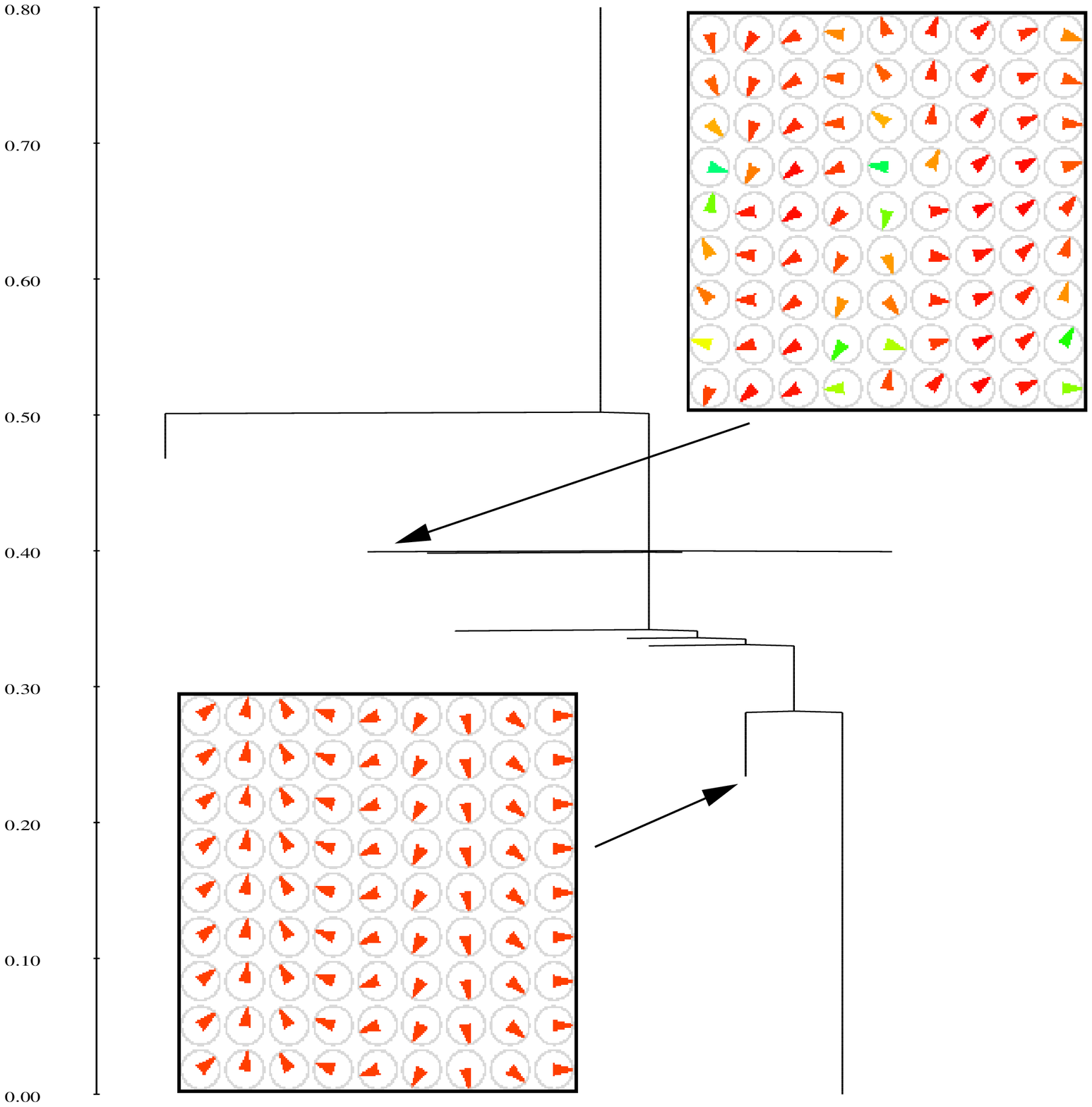}
\includegraphics[width=0.33\textwidth]{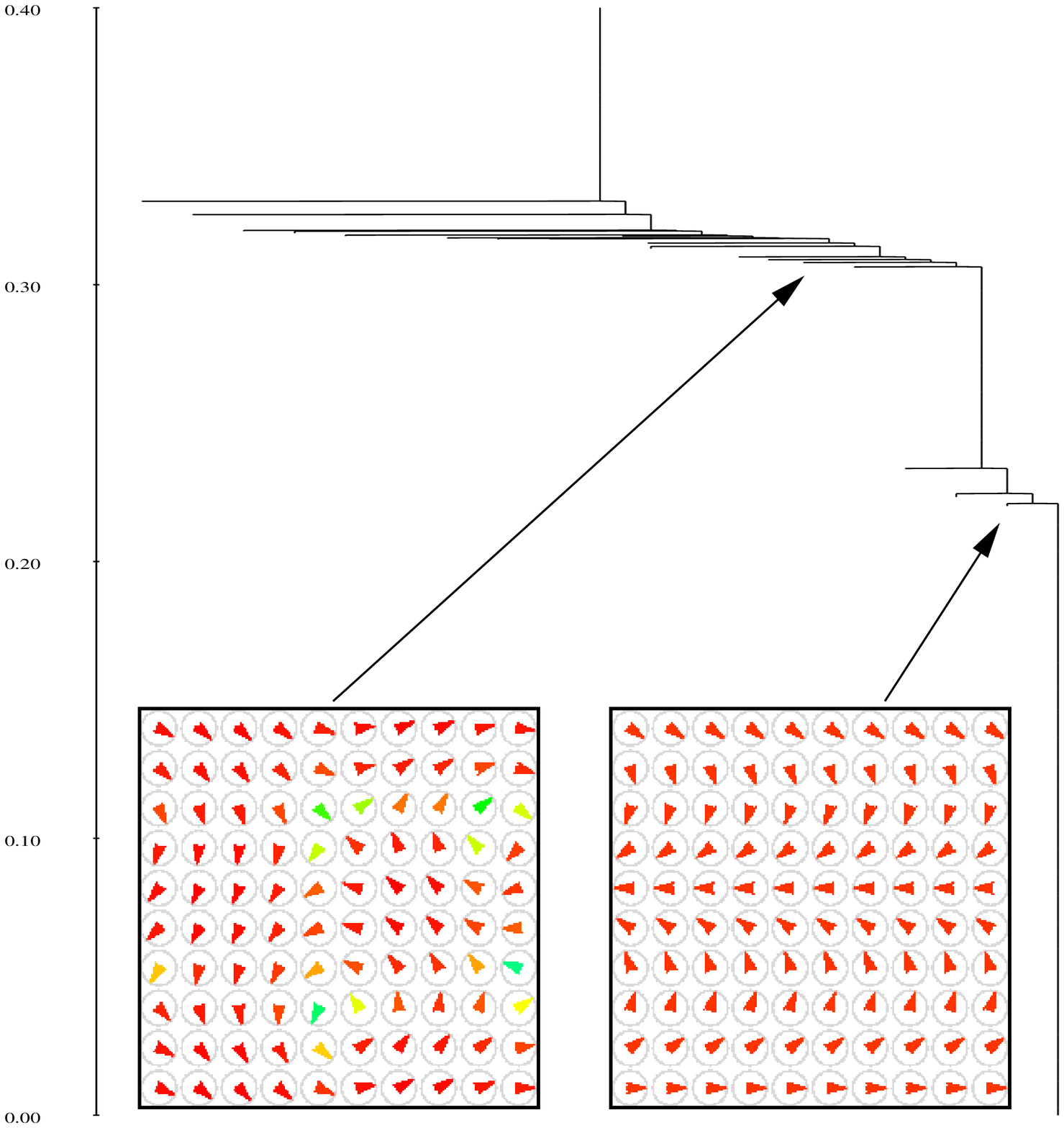}
\caption{Disconnectivity graphs for $5\times 5$, $6\times 6$,
$7\times 7$, $8\times 8$,  $9\times 9$ and  $10 \times 10$ PBC lattices.
 Each of the two insets
represents an example minimum for the corresponding
$N\times N$ lattice. Each arrow in these insets represents the corresponding value of
$\theta_{{\bf i}}$ at the lattice-site ${\bf i}$. At each lattice site ${\bf i}$, we compute the
local energy $\sum_{j=1}^{d}(1- \cos(\theta_{\textbf{i}+\hat{\boldsymbol\mu}_j}-\theta_{\textbf{i}}))$,
which is in
the range $[0,4]$. We colour the arrows
red-orange-yellow-green-blue-indigo-violet from the lowest to highest
local energies.}
\label{fig:discon_PBC}
\end{figure}

In the current contribution we have characterized both minima and transition states,
which enables us to construct the first disconnectivity graphs
for XY models (Figures \ref{fig:discon_APBC} and \ref{fig:discon_PBC}).
These graphs all correspond to the structure expected for efficient relaxation to
the global minimum over a wide range of temperature (or total energy),
namely the `palm tree' motif \cite{walesmw98}.
Locating low-lying minima for these 2D XY models should therefore be relatively
straightforward: relaxation following the intrinsic dynamics of the system
should lead to the global minimum for temperatures of physical interest.
This is the pattern that we associate with good structure-seeking systems \cite{walesmw98,Wales05,Wales10a,Wales12},
including `magic number' clusters such as buckminsterfullerene,
self-assembling mesoscopic structures such as virus capsids,
crystallisation, and
proteins that fold into functional native states on in vivo time scales.

\subsection{Energy Differences}
\label{sec:ediff}

\begin{figure}
\center
\psfrag{N}[tc][tc]{$N$}
\psfrag{EnergyDiff}[bc][bc]{Energy Difference}
\psfrag{0.0}[cr][cr]{0.0}
\psfrag{0.1}[cr][cr]{0.1}
\psfrag{0.2}[cr][cr]{0.2}
\psfrag{0.3}[cr][cr]{0.3}
\psfrag{0.4}[cr][cr]{0.4}
\psfrag{0.5}[cr][cr]{0.5}
\psfrag{0.6}[cr][cr]{0.6}
\psfrag{0.7}[cr][cr]{0.7}
\psfrag{0.8}[cr][cr]{0.8}
\psfrag{4}[tc][tc]{4}
\psfrag{5}[tc][tc]{5}
\psfrag{6}[tc][tc]{6}
\psfrag{7}[tc][tc]{7}
\psfrag{8}[tc][tc]{8}
\psfrag{9}[tc][tc]{9}
\psfrag{10}[tc][tc]{10}
\psfrag{12}[tc][tc]{12}
\psfrag{14}[tc][tc]{14}
\psfrag{16}[tc][tc]{16}
\psfrag{dE1}[cr][cr][0.8]{$dE^{0}_{1,0}$}
\psfrag{dE2}[cr][cr][0.8]{$dE^{0}_{2,1}$}
\psfrag{dE3}[cr][cr][0.8]{$dE^{0}_{3,2}$}
\psfrag{dE4}[cr][cr][0.8]{$dE^{0}_{4,3}$}
\psfrag{dE5}[cr][cr][0.8]{$dE^{0}_{5,4}$}
{\includegraphics[width=0.47\textwidth]{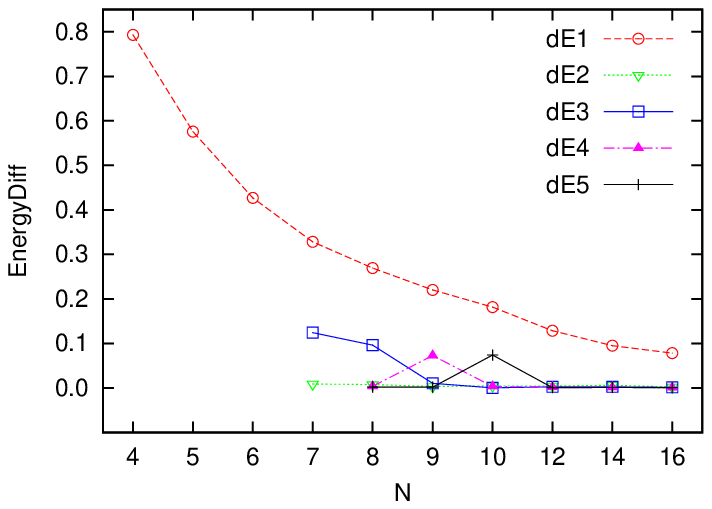}}
{\includegraphics[width=0.47\textwidth]{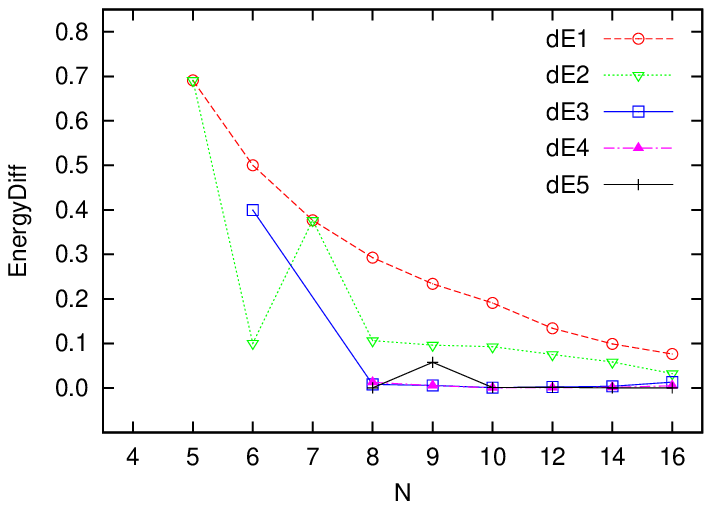}}
\caption{Sequential energy differences between minima when ranked
energetically for APBC (left) and PBC (right) as a function of lattice
dimension $N$. $dE^{0}_{k+1,k}$ is the energy difference between minima $k+1$ and $k$
when arranged in increasing order from $k=0$.
}
\label{fig:energy_diff}
\end{figure}

\begin{figure}
\center
\psfrag{N}[tc][tc]{$N$}
\psfrag{EnergyDiff}[bc][bc]{Energy Difference}
\psfrag{0.0}[cr][cr]{0.00}
\psfrag{0.05}[cr][cr]{0.05}
\psfrag{0.1}[cr][cr]{0.10}
\psfrag{0.15}[cr][cr]{0.15}
\psfrag{0.2}[cr][cr]{0.20}
\psfrag{0.25}[cr][cr]{0.25}
\psfrag{0.3}[cr][cr]{0.30}
\psfrag{0.35}[cr][cr]{0.35}
\psfrag{0.4}[cr][cr]{0.40}
\psfrag{0.5}[cr][cr]{0.50}
\psfrag{0.6}[cr][cr]{0.60}
\psfrag{0.71}[cr][cr]{0.70}
\psfrag{4}[tc][tc]{4}
\psfrag{5}[tc][tc]{5}
\psfrag{6}[tc][tc]{6}
\psfrag{7}[tc][tc]{7}
\psfrag{8}[tc][tc]{8}
\psfrag{9}[tc][tc]{9}
\psfrag{10}[tc][tc]{10}
\psfrag{dE1}[cr][cr][0.8]{$dE^{0}_{1,0}$}
\psfrag{dE2}[cr][cr][0.8]{$dE^{0}_{2,1}$}
\psfrag{dE3}[cr][cr][0.8]{$dE^{0}_{3,2}$}
\psfrag{dE4}[cr][cr][0.8]{$dE^{0}_{4,3}$}
\psfrag{dE5}[cr][cr][0.8]{$dE^{0}_{5,4}$}
{\includegraphics[width=0.47\textwidth]{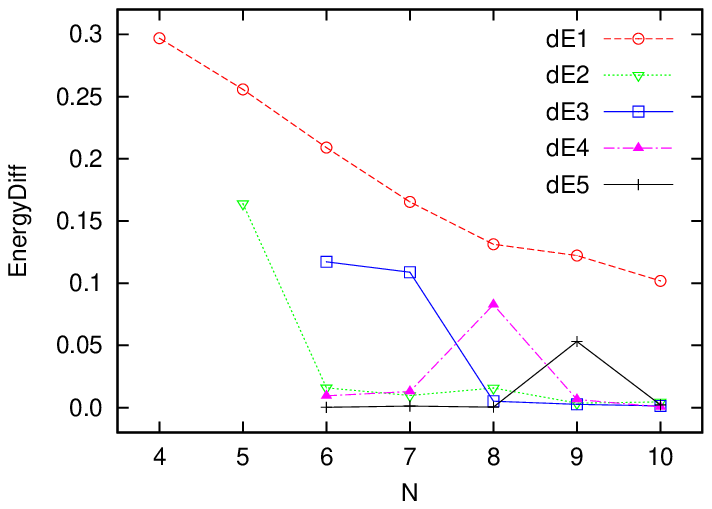}}
{\includegraphics[width=0.47\textwidth]{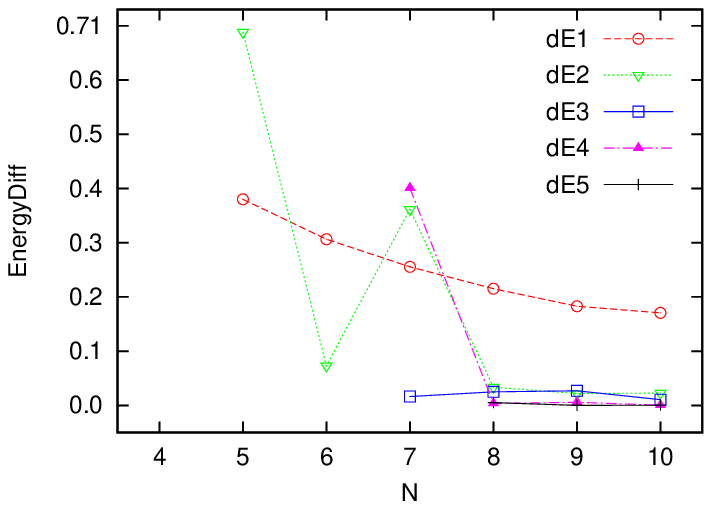}}
\caption{Sequential energy differences between transition states
$k+1$ and $k$ when ranked
energetically for APBC (left) and PBC (right) as a function of lattice dimension $N$.
$dE^{1}_{k+1,k}$ is the energy difference between transition states $k+1$ and $k$
when arranged in increasing order from $k=0$.
}
\label{fig:energy_diff_ts}
\end{figure}

Experimentally, it is not the absolute value of the energy but rather energy differences
that are measured. For the
2D XY model with no disorder, we can in principle study $dE^{i}_{k,l} =
E^{i}_{k} - E^{i}_{l}$, where $E^{i}_{k}$ is the energy of the $k$-th
index $i$ saddle in order of increasing energy.
For the 2D XY model, the global minimum has energy $E^{0}_0 \equiv E_{0}=0$ yielding $dE^{0}_{k,0} = E^{0}_k$.
The sequential energy differences $dE^{i}_{k+1,k}$ are particularly
interesting, since all other
energy differences can be obtained from them. We plot $N$ vs $dE^{0}_{k+1,k}$ in Figure
\ref{fig:energy_diff} and $N$ vs $dE^{1}_{k+1,k}$ in Figure
\ref{fig:energy_diff_ts}. We find that that the sequential
energy differences decay towards zero as $N$ increases.

This observation is similar to results for the 1D XY model with PBC. There, in the
continuum limit, the energies of the local minima are distributed continuously
over the range $[0,1]$. However, there is a spike in the density of
minima at $E_{0}=0$, while other energy values are two-fold degenerate. For small $N$, it appears as
if $dE^{0}_{k,0}$ is decaying towards zero. However, as $N$ becomes large enough,
$dE^{0}_{k,0}$ eventually starts to fill in $[0,1]$, with
a maximum in the
density of minima that approaches the global minimum \cite{Hughes:2012hg}.
In fact, it is the energy
difference between sequential minima, $dE^{0}_{k+1,k}$, that decays towards zero.
For every saddle in
1D with PBC, we can build a higher dimensional analogue that has the same
energy \cite{Hughes:2012hg}. Hence, the energy spectrum of the 2D XY
model with PBC and no disorder
contains at least one copy of the 1D XY model with PBC.

$dE^{0}_{k,0}$ has been
studied in reference \cite{Hughes:2012hg}, where it was found to
decay to zero for the smaller lattice sizes.
The results in Figures \ref{fig:energy_diff} and \ref{fig:energy_diff_ts},
coupled to the previous 1D PBC observations, suggest that
that $dE^{0}_{k+1,k}$ tends to zero, while
$dE^{0}_{k,0}$ fills up a continuous spectrum spanning at least
$[0,1]$. The 2D XY model with APBC seems to follow a similar
pattern to the PBC case.

\subsection{Barrier Heights}
\label{sec:barriers}

\begin{figure}[hp]
\center
\psfrag{N}[tc][tc]{$N$}
\psfrag{barrier}[bc][bc]{Average Barrier}
\psfrag{0.0}[cr][cr]{0.0}
\psfrag{0.2}[cr][cr]{0.2}
\psfrag{0.4}[cr][cr]{0.4}
\psfrag{0.6}[cr][cr]{0.6}
\psfrag{0.7}[cr][cr]{0.7}
\psfrag{0.8}[cr][cr]{0.8}
\psfrag{1.0}[cr][cr]{1.0}
\psfrag{4}[tc][tc]{4}
\psfrag{5}[tc][tc]{5}
\psfrag{6}[tc][tc]{6}
\psfrag{7}[tc][tc]{7}
\psfrag{8}[tc][tc]{8}
\psfrag{9}[tc][tc]{9}
\psfrag{10}[tc][tc]{10}
\psfrag{uphill1}[cr][cr][0.8]{Uphill $\left< \Delta \right>$}
\psfrag{uphill2}[cr][cr][0.8]{Uphill $\left< \Lambda \right>$}
\psfrag{downhill1}[cr][cr][0.8]{Downhill $\left< \Delta \right>$}
\psfrag{downhill2}[cr][cr][0.8]{Downhill $\left<\Lambda \right>$}
\includegraphics[width=0.47\textwidth]{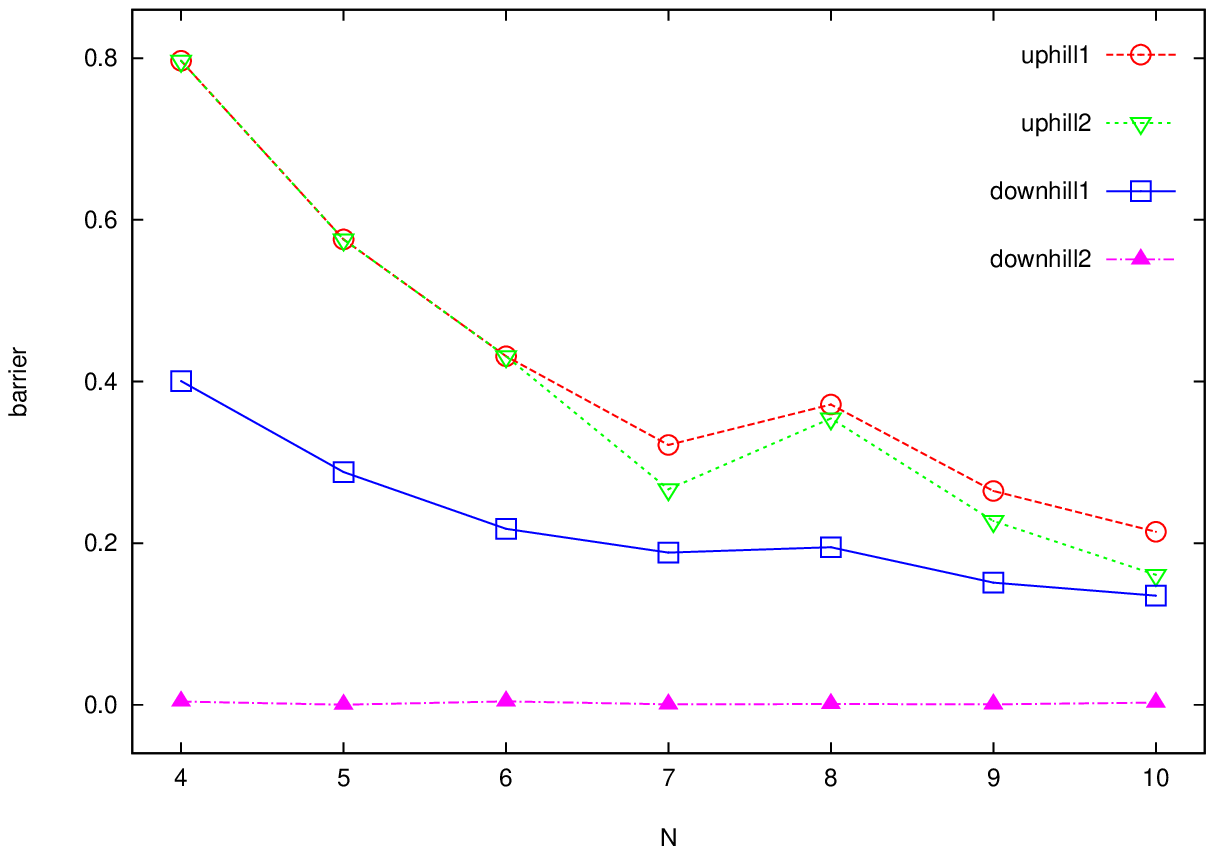}
\includegraphics[width=0.47\textwidth]{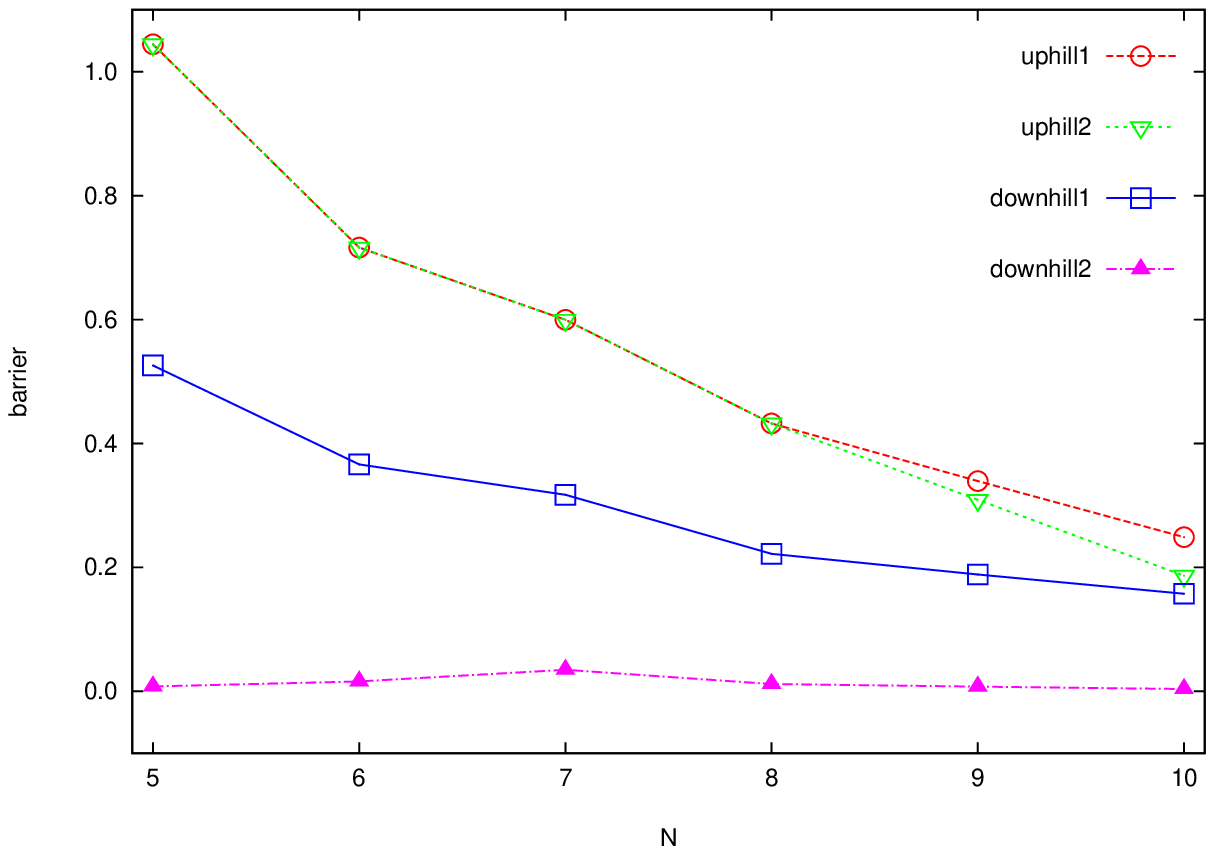}
\caption{Average value of barrier from $ \left< \Delta \right>$ and $\left< \Lambda \right> $ for APBC and PBC.}
\label{fig:barrier_heights}
\end{figure}

The average uphill/downhill barrier between minima and transition states can be defined \cite{WalesDoye:2002} as
\begin{align*}
\left< \Delta \right> & = \left< E_{\text{ts}} \right> - \Sigma_\gamma n^\gamma_{\text{ts}} E^\gamma_{\text{min}} /2n_{\text{ts}}
\end{align*}
where $E^\gamma_{\text{min}}$ is the energy of the minimum $\gamma$ and $n^\gamma_{\text{ts}}$
is the number of transition states connected to that minimum. The naive uphill/downhill barrier is given by
\begin{align*}
\left< \Lambda \right> & = \left< E_{\text{ts}} - E_{\text{min}} \right>.
\end{align*}
As noted in \cite{WalesDoye:2002}, the average over minima in the second term
of $\left< \Delta \right>$ is usually weighted towards the lower energy minima,
since they are connected to more transition states. This organisation makes
$\left< \Delta \right>$ larger than $\left< \Lambda \right>$, as we
see in the plots of the average barriers in Figures
\ref{fig:barrier_heights}. In these plots, only distinct non-degenerate
lowest-energy rearrangements were considered in the averages.

%for non-degenerate and degenerate rearrangements, respectively.
%Here, a degenerate rearrangement corresponds to a pathway connecting
%symmetry-related minima (or the same minimum), which must have the same
%potential energy \cite{Wales:04}.
%The distribution is shown in Figure \ref{fig:hist_barrier_heights}.

% {\bf{Here}}I don't know how to interpret these figures
% yet. I think these are interesting quantities, but I am still thinking what
% information it gives for the XY model. On the other hand, since ours is the
% first ever work on full-fledgedly exploring the PEL of the 2D XY model, we may
% still want to put this as an `observation' even if there is no particularly
% helpful interpretation of them we can think of for now.

\section{Conclusions}
\label{sec:conclusions}

The potential energy landscape has been examined for the two-dimensional XY
model (with no disorder) with both periodic and anti-periodic boundary conditions.
Lattices with up to $N=10$ lattice sites in each direction, i.e.~100 spins, were considered, focusing on the
potential energy distribution of minima and the transition states (saddles of index one)
that connect them.
As expected \cite{stillingerw84,2003JChPh.11912409W}, the number of stationary points
increases roughly exponentially with the number of degrees of freedom.
Knowledge of the pathways that connect the local minima enables us to construct
the first disconnectivity graphs for the XY model, and hence visualize the potential
energy landscape.
These graphs reveal that the landscape is funnelled in each case, with a well-defined
global minimum, and small downhill barriers connecting this structure to the higher-energy
configurations.
Hence all of these 2D XY landscapes belong to the class of systems identified as good `structure seekers',
which includes `magic number' atomic and molecular clusters, naturally occurring proteins,
and self-assembling mesoscopic systems, including crystals \cite{walesmw98,Wales05,Wales10a,Wales12}.
Minimization from random starting points confirms that the global minimum is readily
located in each case; the funnelled organisation of the landscape is reinforced by
the existence of relatively large basins of attraction for the global minima compared to the higher energy minima,
and this effect grows with increasing lattice size.

Although the samples of stationary points are not exhaustive for the larger lattice sizes,
we can draw some further general conclusions.
First, for a given lattice size, $N$, there are more minima for antiperiodic boundaries than
for periodic boundary conditions.
Second, as $N$ increases the energy range spanned by the local minima increases,
as one might expect from extensivity of the energy.
This effect is also visible in the disconnectivity graphs.
However, the probability distribution for the energy of the local minima
tends to shift towards the global minimum for larger lattice sizes.

The trends we have identified have far-reaching implications for the thermodynamics
and global kinetics of the 2D XY model, which we will investigate in future work.
Given the wide-ranging applications of this model, which include superconductivity,
superfluidity, liquid crystals, Josephson junctions, and the fundamental
importance of this Hamiltonian in lattice gauge theory \cite{Maas:2011se,Mehta:2009,Mehta:2010pe},
the energy landscape perspective may provide new insight into a variety of different
research fields.

% Conclusions so far:
% \begin{enumerate}
% \item \textit{Is it true?} In both PBC and APBC cases, the differences
%   between two adjacent levels decreases as the lattice size
%   increases. (\textbf{Ciaran}: I think the
%   energy spectrum becomes continuous (in fact, for the trivial case,
%   the 2D PBC has to have a copy of the 1D case, and the 1D case has
%   spectrum [0,1]). I think the differences between minima become
%   smaller, but that doesn't mean they all become degenerate. We
%   cannot build a 2D APBC minimum from a 1D one, so a possible
%   interesting thing is that the 2D APBC case is different from 1D
%   APBC,  but behaving similar to 2D PBC from our results.)
% \item As $N$ increases, the energies of the
%   minima appear to concentrate near the global minimum. (This
%   looks like Zwanziger's conjecture). {\bf{Ciaran:}} This is seen in 1D
%   PBC. The density of states at the global minimum has a spike, and
%   other minima are two-fold degenerate. The same argument about 2D PBC
%   containing 1D PBC applies here, so not surprising 2D PBC contains
%   information similar to 1D PBC. But 2D APBC behaves different from
%   1D APBC but similar to 2D PBC.
% \end{enumerate}

\bigskip
DM was supported by the U.S. Department of Energy under contract no.
DE-FG02- 85ER40237 and DARPA Young Faculty Award. CH acknowledges support from Science and Technology Facilities Council and the Cambridge Home and European Scholarship
Scheme. MS acknowledges support by the Research Executive
Agency (REA) of the European Union under Grant Agreement
PITN-GA-2009-238353 (ITN STRONGnet).
DJW gratefully acknowledges support from the EPSRC and the ERC.

%\bibliographystyle{unsrt}
%\bibliography{bibliography_NPHC_NAG}

\newpage

\section*{Tables}

\vfill
%\begin{widetext}
\begin{table*}[htbp]
\begin{tabular}{|l||c|c|c|c|c|c|c|c|c|c|c|}
\hline
\multicolumn{1}{|c||}{$N$} & 3 & 4 & 5 & 6 & 7 & 8 & 9 & 10 & 12 & 14 & 16 \\\hline\hline
minima (APBC) & 2 & 66 & 202 & 146 & 1570 & 7170 & 24626 & 99207 & 849329 & 2826736 & 5606875 \\ \hline
distinct minima (APBC) & 1 & 2 & 2 & 2 & 4 & 13 & 20 & 49 & 298  & 1671  & 10876  \\ \hline
saddles of index 1 (APBC) & 18 & 288 & 850 & 3864 &  13890 &
27456  & 51234  & 52572  & \multicolumn{3}{c|} {\multirow{3}{*}{}} \\ \cline{1-9}
distinct saddles of index 1 (APBC) & 1 &2 &3&
 8& 15 &73 &201& 615 & \multicolumn{3}{c|} {\multirow{3}{*}{}} \\ \hline
minima (PBC) & 1 & 1 & 9 & 27 & 9  & 681  & 44000  & 13918 & 111699 & 704547 & 2593377  \\ \hline
distinct minima (PBC)   & 1 & 1 & 3 & 4 & 3 & 8 & 14 & 44 & 257 & 2266 & 23352 \\ \hline
saddles of index 1 (PBC) & 9 & 16  & 71 & 234 &  277 &  2540 &
3587  & 3854 & \multicolumn{3}{c|} {\multirow{3}{*}{}} \\ \cline{1-9}
distinct saddles of index 1 (PBC) & 1 &1 &3&
5&3&17&30 & 115 & \multicolumn{3}{c|} {\multirow{3}{*}{}} \\ \hline
\end{tabular}
\caption{The number of minima and saddles of index one located for different lattice sizes
$N\times N$, with both PBC and APBC.}
\label{tab:minima-transition-states}
\end{table*}
%\end{widetext}
\vfill

\newpage

\section*{Figures}

\vfill
\vfill

\vfill
\vfill

\newpage
\vfill
\vfill

\newpage

\vfill
\vfill

\newpage

\vfill
\vfill

\newpage

\vfill
\vfill

\end{document}